# Growth and Strain Relaxation Mechanisms of InAs/InP/GaAsSb Core-Dual-Shell Nanowires

Omer Arif,[§] Valentina Zannier,[§]* Ang Li,[#] Francesca Rossi,[∥] Daniele Ercolani,[§] Fabio Beltram,[§] and Lucia Sorba[§]

[§] NEST, Istituto Nanoscienze-CNR and Scuola Normale Superiore, Piazza San Silvestro 12, I-56127 Pisa, Italy.

[#] Beijing Key Laboratory of Microstructure and Properties of Solids, Beijing University of Technology, 100124 Beijing, China.

[∥] IMEM-CNR, Parco Area delle Scienze 37/A, I-43124 Parma, Italy.

KEYWORDS: nanowires, radial heterostructures, strain relaxation, geometric phase analysis.

ABSTRACT: The combination of core/shell geometry and band gap engineering in nanowire heterostructures can be employed to realize systems with novel transport and optical properties. Here, we report on the growth of InAs/InP/GaAsSb core-dual-shell nanowires by catalyst-free chemical beam epitaxy on Si (111) substrates. Detailed morphological, structural and compositional analysis of the nanowires as a function of growth parameters were carried out by scanning and transmission electron microscopy, and by energy-dispersive X-ray spectroscopy.

Furthermore, by combining the scanning transmission electron microscopy-Moiré technique with geometric phase analysis, we studied the residual strain and the relaxation mechanisms in this system. We found that InP shell facets are well developed along all the crystallographic directions only when the nominal thickness is above 1 nm, suggesting an island-growth mode. Moreover, the crystallographic analysis indicates that both InP and GaAsSb shells grow almost coherently to the InAs core along the <112> direction and elastically compressed along the <110> direction. For InP shell thickness above 8 nm, some dislocations and roughening occur at the interfaces. This study provides useful general guidelines for the fabrication of high quality devices based on these core-dual-shell nanowires.

INTRODUCTION

Semiconductors nanowires (NWs) can be used as building blocks for electronic, photonic devices and sensors due to their unique properties, such as high surface-to-volume ratio, free-standing nature and capability to relax elastic stress in two-dimensions.[1] One of the most appealing NW feature is the possibility to create heterostructures by combining different materials in both axial and radial directions.[2,3] In recent years, researchers are paying much attention on the so-called core-shell (CS) NWs because radial geometry can improve the performance and/or add new properties in the devices.[4,5] For example, surface passivation by the introduction of one or more shells around the core can enhance the radiative emission efficiency reducing the carriers surface recombination.[6] Many core-shell NWs based on III-V semiconductors were demonstrated: InAs/InP,[7] InAs/GaAs,[8] InAs(Sb)/GaSb,[6] InAs/GaSb,[9] and GaAs/GaSb.[10] Among these, InAs/GaSb core-shell NWs attracted great attention because of their peculiar properties that provide a useful platform for many applications. As a matter of fact, both InAs and GaSb have very small effective masses with high electron and hole mobility, respectively. Moreover, they have a type-III broken gap band alignment[11] and a very low lattice mismatch of 0.6%.[12] All these properties, make InAs/GaSb core-shell NWs suitable for applications in devices like tunnel field effect transistors,[13] Esaki diodes,[11] frequency multipliers,[6] and for fundamental studies on spin states[14] and electron-hole hybridization.[15] Indeed, electronic devices fabricated with these heterostructures implement radial interface between n-type and p-type conductors, and can display negative differential resistance owing to transport across the broken-gap junction.[11] Further interesting electronic device configurations can be achieved if the carriers are separated in two channels, i.e in the InAs core and GaSb shell respectively. To this end, in the present work we inserted thin InP barriers of different thickness in between InAs core and GaSb shell. We choose InP as barrier material because of its small lattice mismatch with InAs and GaSb, and its larger band gap compared to both InAs and GaSb. The InP barrier will provide a separation of the carriers in two distinct channels, i.e. electrons in the InAs core and holes in the GaSb shell and it will allow to realize novel electronic devices.

It is well known that electronic and optical properties of semiconductor heterostructures are affected by the presence of strain fields arising from lattice mismatch between the combined materials.[16,17] Pseudomorphic growth of NWs in a core-shell geometry implies a coherency limit in both diameter of core and shell thickness, that depends on the lattice mismatch between the two materials.[18,19,20] For a given NW core diameter, the shell material will grow coherently strained only below a critical thickness, while above this it is energetically favored to produce misfit dislocations that degrade device performance.[21] The type, density and distribution of dislocations strongly rely on the material system.[22,23] For example, Treu *et. al.* reported on InAs-InAsP core-shell NWs coherently grown with 10 nm thick InAsP shell with P content of 9%, but a shell of InP of the same thickness showed dislocations. The InAsP shell around the InAs core enhanced photoluminescence for low P content, but drastically decreased light emission for higher P contents. Therefore, it is clear that a fundamental step towards the fabrication of high performance devices with heterostructured NWs is the investigation of strain relaxation mechanisms and critical shell dimensions.

In this work, we studied the catalyst-free growth of InAs/InP/GaSb core-dual-shell (CDS) NWs on Si (111) substrates by chemical beam epitaxy (CBE). In contrast to the well-known Au-catalyzed NW growth method, the catalyst-free approach is preferred for CDS growth because there is no metal particle on the top of the NWs so that the radial growth can be enhanced compared to the axial growth.[4] The influence of growth temperature on the InP shell morphology, crystal structure and chemical composition was investigated by scanning electron microscopy (SEM), energy dispersive X-ray spectroscopy (EDX) and high-resolution transmission electron microscopy (HR-TEM). We investigated also the growth of InAs/GaSb CS NWs and we found that there is a significant As incorporation in the GaSb shell, so the actual composition of the shell is $GaAs_{0.4}Sb_{0.6}$. Finally, the growth of both InP and GaAsSb shells of well-defined thickness around the InAs core was achieved. We analyzed the strain relaxation in these InAs/InP/GaAsSb CDS NWs as a function of the InP shell thickness with the help of scanning transmission electron microscopy Moiré pattern (STEM- Moiré) and combination of high-resolution STEM (HR-STEM) imaging and strain mapping by geometric phase analysis (GPA). To the best of our knowledge, this type of NW system was not reported before, and we believe that this work provides useful information for the realization of NW-based devices with spatially separated charge carriers, which may also open the way to the study of electron-hole pairing effects and Coulomb drag phenomena."

EXPERIMENTAL DETAILS

Catalyst-free InAs/InP/GaAsSb CDS NWs were grown on Si (111) substrates by using chemical beam epitaxy (CBE) in a Riber Compact-21 system. The metalorganic (MO) precursors used for the NW growth are trimethylindium (TMIn), tertiarybutylarsine (TBAs), tertiarybutylphosphine (TBP), triethylgallium (TEGa), and trimethylantimony (TMSb).

In the first step, InAs core NWs were grown on Si (111) substrates by the vapor-solid (VS) method. Substrate preparation and growth procedure are reported in Ref 24: prior to the growth, the Si (111) substrate was annealed at 700 ± 10 °C under TBAs flow for 15 min and then cooled down to 300 ± 10 °C to start the growth of InAs NWs by opening MO lines with pressures of 0.3 Torr and 3.3 Torr of TMIn and TBAs, respectively. This step is crucial for the nucleation of the NWs. After nucleation the temperature was ramped up to 430 ± 10 °C in 10 min and finally the growth was continued for two hours with MO line pressure of 0.2 Torr and 3.3 Torr of TMIn and TBAs to obtain NWs with the desired aspect ratio.[24] For the growth of the InP shell around the InAs core NWs, the temperature was decreased from 430 ± 10 °C to the 350-380 °C range under the TBAs flux at the end of InAs growth, and then InP growth was immediately started by using TMIn and TBP line pressures of 0.3 Torr and 1 Torr, respectively. After optimization of the growth temperature, a series of samples with different InP-shell thickness was grown. Finally, the outer GaSb shell was grown in a second growth step: the InAs/InP NWs were kept in UHV environment at room temperature for 30 minutes, while the growth chamber was pumped in order to decrease As and P residual background. Then the sample was re-introduced into the growth chamber and warmed up under TMSb pressure to 370 ± 10°C, at which we started the shell growth using TEGa and TMSb line pressures of 0.50 Torr and 0.43 Torr, respectively. The growth time of the GaSb shell was varied to obtain different thicknesses. Finally, the growth was terminated by cooling the sample under TMSb flux.

NW morphology was characterized by scanning electron microscopy (SEM) in a Zeiss Merlin field emission microscope operated at 5 KeV by acquiring top- and tilted-view (45°) images. The crystal structure, elemental composition and shell thickness of NW were measured with transmission electron microscopy (TEM) with a JEOL JEM-2200FS microscope operated at 200 keV, equipped with an in-column Ω filter and a detector for X-ray energy-dispersive spectroscopy (EDX). Imaging was performed either in HR-TEM mode combined with zero-loss energy filtering or scanning (STEM) mode using a high-angle annular dark-field (HAADF) detector yielding atomic-number (Z) contrast. For TEM observation, the NWs were mechanically transferred to carbon-coated copper grids.

For strain analysis, the as-grown NWs were transferred to a suitable substrate and cross-sectional STEM lamellae were cut by Focused Ion Beam (FEI-Helios 650) with a Pt deposited protection layer. The STEM-Moiré pattern of the samples was generated by a probe aberration-corrected STEM (FEI-Titan-Themis) operated at 300 kV equipped with a monochromator. The HAADF detector distance was chosen to gain a collection angle ranging from 40 mrad to 200 mrad, and the scanning resolution was set as 1024 by 1024 pixels with a pixel resolution of 0.19 nm. In order to avoid artifacts due to sample drift, the dwell time of each scanning point was set as 16 μs (18.23 s per frame). The Geometric Phase Analysis (GPA) to calculate the lattice strain was performed by STEM_Cell software.[25]

RESULTS AND DISCUSSION

Figure 1 (a) shows the InAs NW core (top and tilted SEM images). The average length and diameter are 1.5 ± 0.1 μm and 130 ± 5 nm, respectively. From the top-view image we can observe that the NWs have a hexagonal cross-section with a flat top facet parallel to the (111) substrate surface. The tilted-view SEM image shows that the NWs do not exhibit detectable tapering displaying a uniform diameter along the entire length.

In order to investigate the influence of growth temperature on the InP shell, we grew a series of samples in which all other parameters (line pressures and growth time) were kept constant. Figure 1 (b)-(d) shows the InAs/InP radial heterostructured NWs obtained at different growth temperatures. The InP shell grown at 360 ± 10 °C (panel b) shows very rough sidewalls and the top facet is not flat showing an irregular cross-section. By increasing the growth temperature to 370 ± 10 °C (panel c), we found NWs with regular hexagonal cross-section, smoother sidewalls and flat top facet. A further increase of the growth temperature to 380 ± 10 °C (panel d) gives NWs with still smooth sidewalls, but larger axial InP growth, resulting in a visible InP top segment.

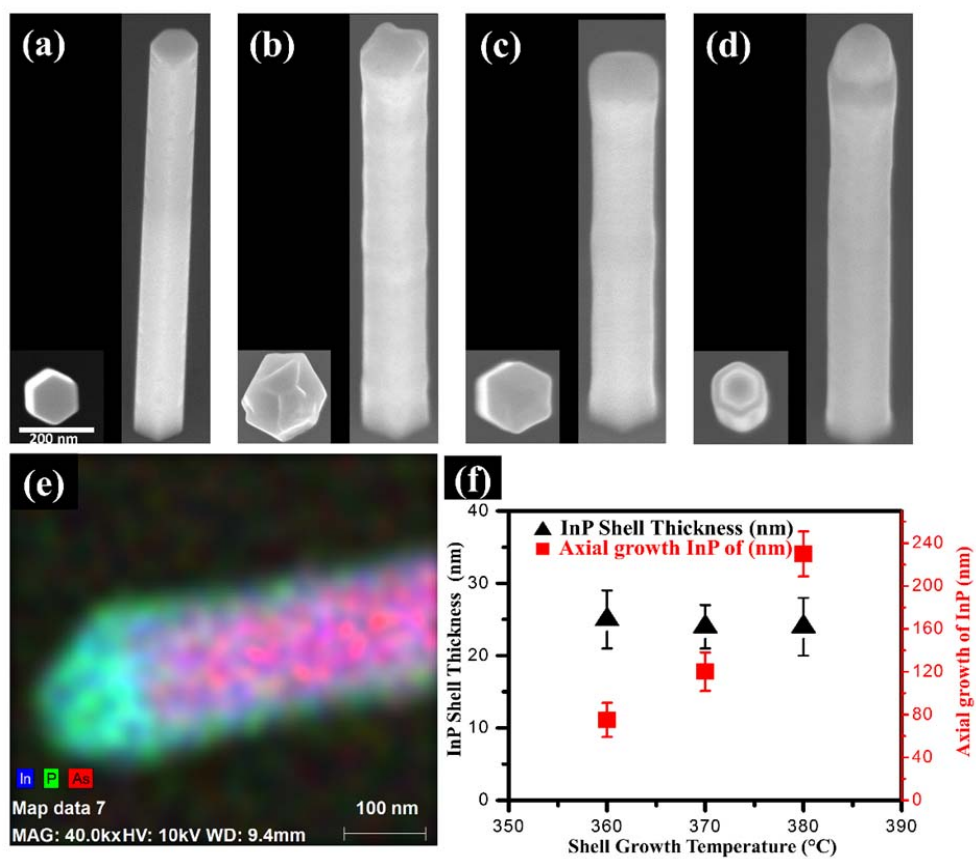

**Figure 1.** (a)-(d) SEM images of catalyst-free InAs/InP CS NWs : top view (left) and 45°-tilted (right) images of InAs core NW (a) and of InAs/InP CS NWs growth at different shell growth temperatures indicated in the panels: (b) 360 °C, (c) 370 °C, and (d) 380 °C. Scale bar is the same in all panels: 200 nm. From the top-view images the hexagonal cross-sections of all NWs are clearly visible. (e) EDX map of the upper portion of the InAs/InP CS NW displayed in panel (d). (f) Plot of InP radial and axial growth versus InP shell growth temperature. The axial and radial InP thickness were measured by EDX line scans.

In order to quantify the thickness of the InP shell and the axial segment grown on the InAs NW tip we acquired EDX compositional maps like the one shown in Fig. 1 (e) for 10 NWs at each growth temperature and from the axial and radial line scans we determined the average values reported in Fig. 1 (f). Data show that the radial growth rate of InP shell is constant in this range of growth temperature explored, while the axial growth rate of InP increases linearly with the temperature. This increased axial growth rate stems from the higher kinetic energy of the In adatoms on the NW sidewalls at higher temperatures: they can more efficiently reach the top facet of the NW and contribute to axial growth. Based on this analysis we selected 370 ± 10 °C as the optimal growth temperature since it is high enough to provide an InP shell with smooth sidewalls, but at the same time low enough to keep a low axial growth rate.

Figure 2 (a) shows the HR-TEM image of a representative InAs/InP CS NW with the InP shell grown at the optimized temperature (370 ± 10 °C) for 3 minutes. Several defects such as stacking faults and twins perpendicular to the growth axis are visible in both InAs core and InP shell. It is known that the crystal structure of InAs NWs obtained by catalyst-free vapor-solid growth is a mixture of zinc blende and wurtzite segments, so that the NWs have several stacking faults perpendicular to the growth direction.[24] Such defects propagate also in the InP shell. The interface between the two materials is visible in the HR-TEM image (highlighted with the red lines). We measured the shell thickness of more than 10 NWs at different positions along the growth axis and we found an average thickness of 4.0 ± 0.5 nm (see a representative STEM-HAADF image in Figure 2 (b)), accordingly with what expected from the growth rate. We couldn't evaluate the chemical composition of this very thin shell from EDX analysis because the P signal is very low and the atomic % quantification can be misleading. A more accurate EDX analysis, allowing the precise element quantification, is performed in cross-sectional lamellae of the final CDS NWs, as it will be shown later.

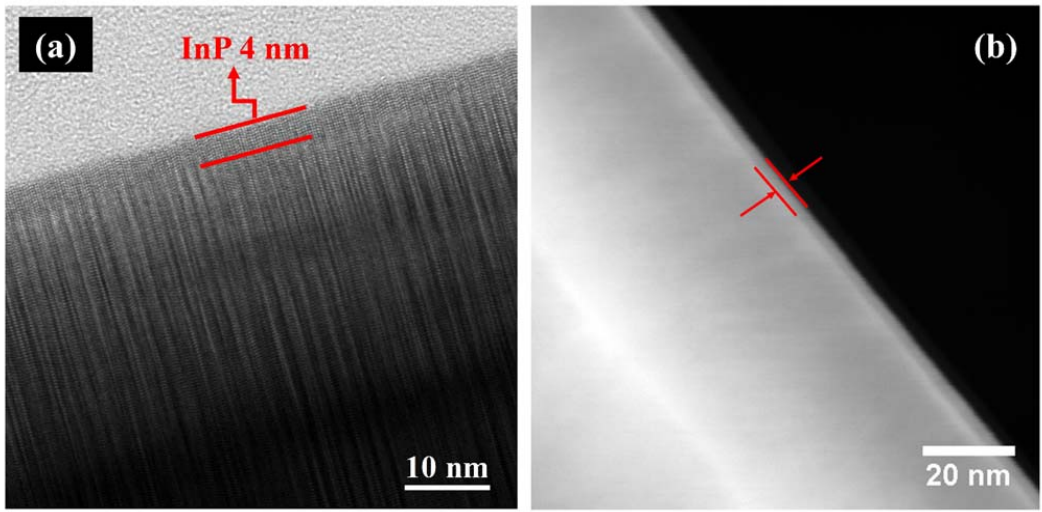

**Figure 2.** (a) Bright field HR-TEM image, acquired in [110] zone axis, of an InAs/InP core shell NW. The red lines indicate the InP shell. (b) STEM-HAADF image, acquired in [112] zone axis, of a NW from the same sample confirming the smooth and homogenously thick (4 nm) InP shell around the InAs core.

In order to optimize the growth of a GaSb shell at the same temperature of the InP shell with the best morphology, which is also known to be a good growth temperature for GaSb,[26] we first investigated GaSb growth directly around the InAs NWs (without InP) at 370 ± 10°C using different precursor line pressures, keeping the same III/V ratio, previously optimized.[11] The best results in terms of growth rate were obtained by using 0.5 Torr and 0.43 Torr of TEGa and TMSb line pressures, respectively. Figure 3 shows the CS NWs obtained after 90 min of GaSb growth. Panel (a) shows a representative STEM-HAADF image of a single NW, aligned along the [112] zone axis. The STEM analysis of various NWs confirms the presence of a shell with uniform thickness all along the growth axis for the whole NW length, around the InAs core.

Panels (b) and (c) show the EDX analysis of the same NW depicted in Fig. 3 (a), performed by using the Kα emissions of Ga and As, and the Lα emissions of In and Sb. The EDX composition map shown in panel (b) clearly shows the presence of a shell of 17 ± 1 nm thickness around the core as indicated with green color. The shell growth rate is therefore 0.21 ± 0.01 nm/min. The growth rate is very low probably because of the low fluxes and growth temperature selected in order to ensure homogenous and smooth sidewalls. Panel (c) shows the cross-sectional line profile of the NW, from which we confirmed the presence of In in the core and Ga in the shell. The apparent presence of In in the shell can be explained considering that In and Sb signals are partially overlapping. The same overlapping influences the intensity of the Sb signal in the core part of the NW. On the other hand, the As signal is well resolved, therefore the non-zero intensity of the As signal in the shell really indicates the presence of As together with Ga and Sb here. The reason of this unintentional As incorporation in the shell is probably the persistence of an As background in the growth chamber during the GaSb deposition, due to the very high As line pressure used for the catalyst-free InAs NW, combined with the very low growth rate of the shell. The precise quantification of the chemical composition of the GaAsSb shell is done in the CDS final structure, as it will be shown later.

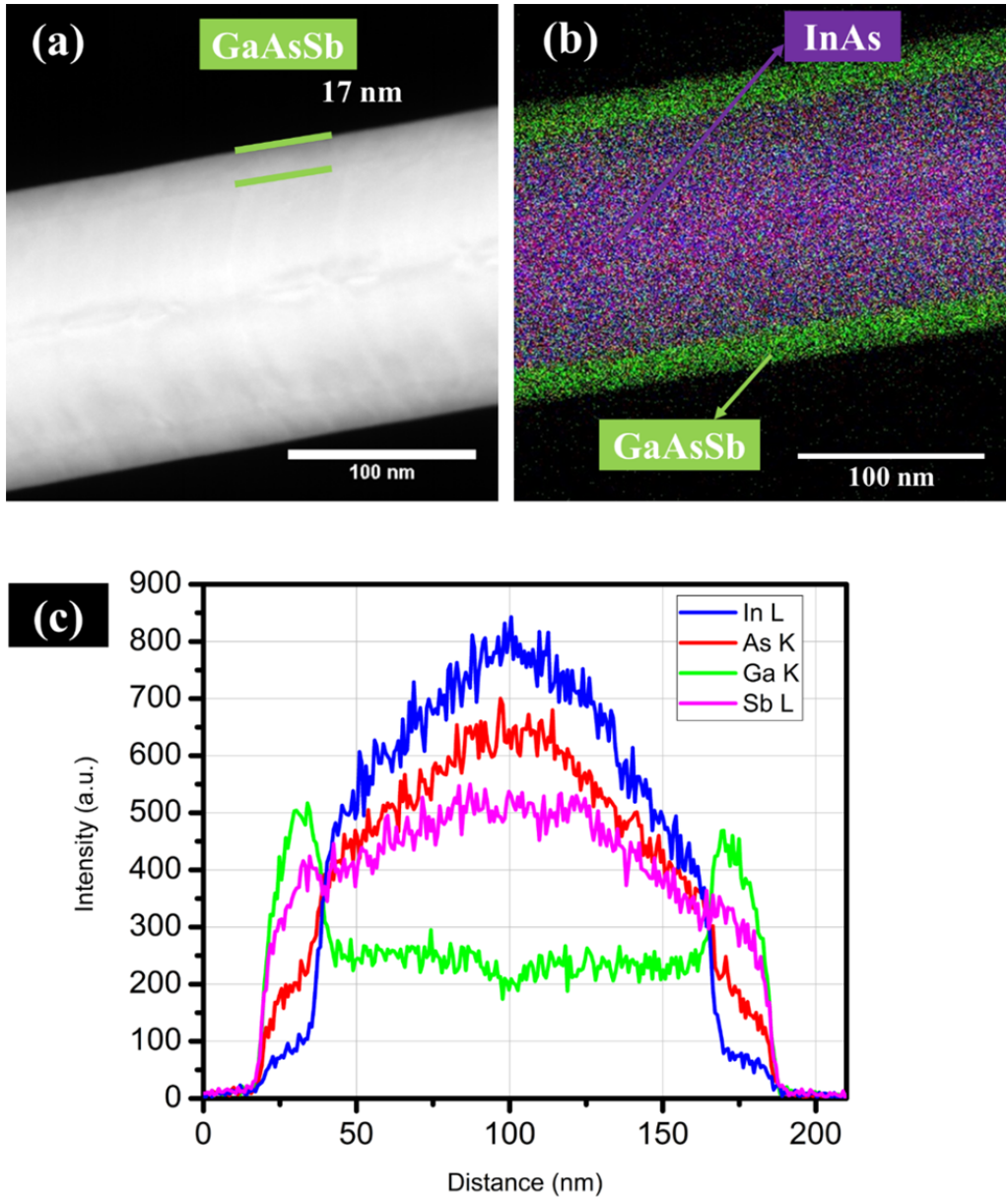

**Figure 3.** (a) STEM-HAADF image of one representative InAs/GaAsSb CS NW obtained after 90 min of GaSb deposition at 370 ± 10°C with TEGa and TMSb fluxes of 0.5 Torr and 0.43 Torr, respectively. (b) EDX compositional map and (c) elemental line profiles in cross-section of the same InAs/ GaAsSb NW depicted in (a). The results indicate the growth a GaAsSb shell of 17 ± 1 nm thickness around InAs core.

After growth optimization of InP and GaAsSb shells separately on the InAs NW core, we combined the three materials in single InAs/InP/GaAsSb CDS heterostructured NWs. Figure 4 shows a STEM-HAADF image of a representative InAs/InP/GaAsSb CDS NW, aligned along the [112] zone axis, with the corresponding compositional EDX map. InP and GaSb growth times were 10 and 140 minutes, respectively, at the same growth temperature of 370 ± 10 °C. The resulting thicknesses of InP and GaAsSb shells measured from STEM images were 14 ± 1 nm and 30 ± 2 nm, respectively. The analysis of several NWs confirmed the presence of the two shells with uniform thicknesses along the growth axis for the whole NW length.

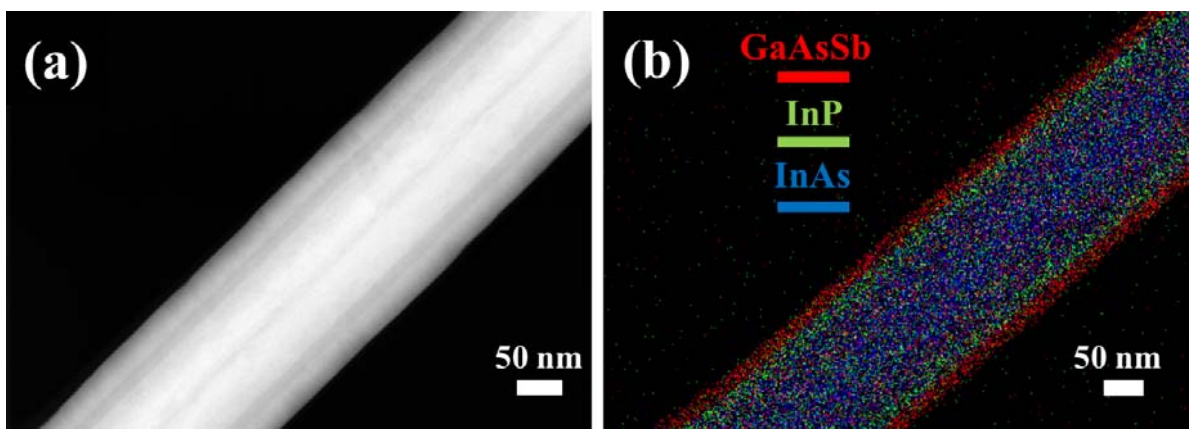

**Figure 4.** (a) STEM image acquired in [112] zone axis and (b) EDX compositional map of the middle region of a InAs/InP/GaAsSb CDS NW grown with flux ratio of TMIn/TBP 0.3/1.0 Torr and growth time of 10 min at 370 ± 10°C for the InP shell (resulting in a 14 ± 1 nm thick InP shell) and flux TEGa/TMSb 0.5/0.43 and growth time 140 min at 370 ± 10°C for the GaSb shell (resulting in a 30 ± 2 nm thick GaAsSb shell).

The HAADF intensity profiles of InAs/InP and InAs/InP/GaAsSb NWs in cross-section along <112> zone axis (shown in Fig. S1 of the supporting information) reveal that the InAs/InP CS NWs have six {110} side facets, so the InP shell grew keeping the same side facet orientation than the InAs core (see panel (a) of Fig. S1). Instead, the InAs/InP/GaAsSb CDS NWs show twelve facets: six belong to the {110} family and six belong to the {112} family of planes. This behavior of the GaAsSb shell developing {112} side facets was already observed in the growth of InAs/GaSb[26] and InAs/GaAs[8] CS NWs, and explained by the Wulff's construction[27] that ascribes the final shape of a crystal to the different surface energies of the different facets. Also in our case, as it will be shown later, the different GaAsSb growth rates in the two directions suggest that the {112} facets have a lower surface energy than the {110} ones, and this explains the development of such facets in the GaAsSb shell.

In order to study the strain accommodation at the heterointerfaces in the CDS NWs, we prepared cross-sections perpendicular to the NW growth direction of three samples having different InP shell nominal thickness (1, 4 and 8 nm) and same GaAsSb shell nominal thickness (12 nm) by focused ion beam (FIB), and these cross-sectional lamellae were inspected by STEM. In order to identify structural defects and distortions, STEM Moiré patterns[28] were acquired for each sample by aligning the scanning direction along the <112> crystallographic direction and by carefully choosing the line resolution during the scanning. In this case Moiré patterns are generated since the scan step is comparable to the lattice periodicity and fringes are formed in one direction, similarly to a single set of crystalline lattice planes. The results are shown in Fig. 5 for the three samples with different InP thickness: 1 nm (a, d, g), 4 nm (b, e, h), and 8 nm (c, f, i). In all samples the different materials can be easily identified thanks to the Z contrast of the HAADF imaging mode, as visible in panels (a), (b) and (c), which are the STEM-Moiré images of the entire lamellae: the inner part corresponds to the InAs core, the dark central ring represents the InP shell and the external ring is the GaAsSb shell.

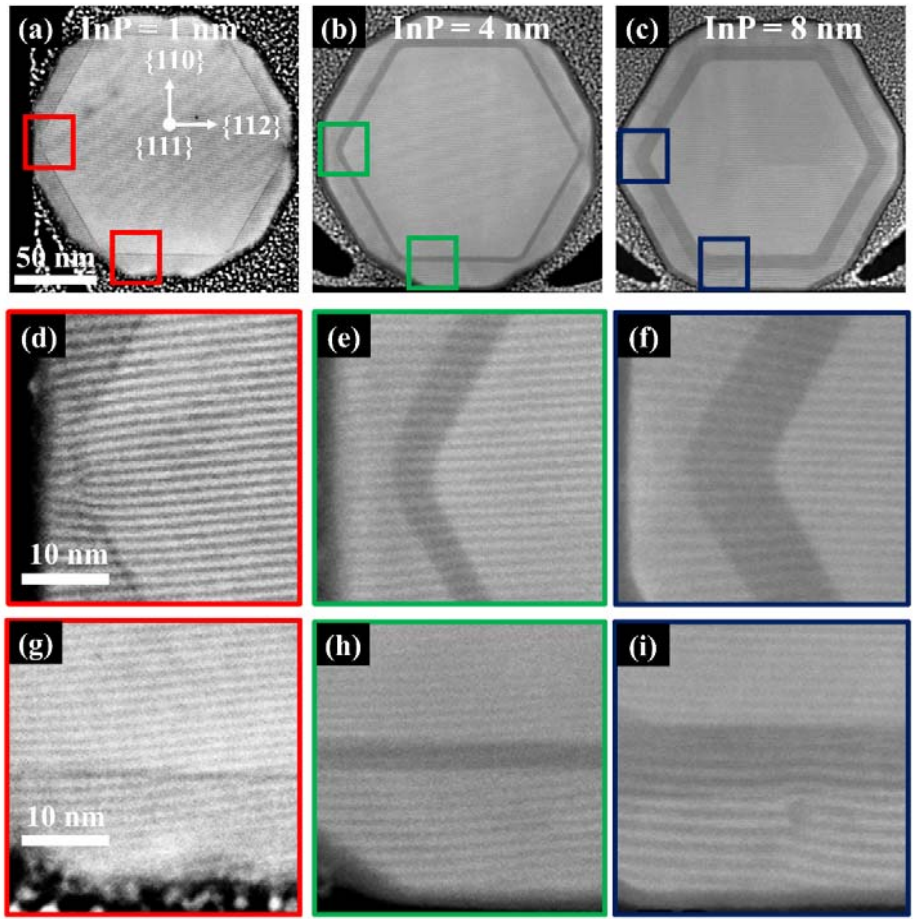

**Figure 5.** (a)-(c). STEM-Moiré images of entire cross-sectional lamellae of the three samples having different InP shell thickness (indicated in the panels). (d)-(f) STEM-Moiré patterns of the selected regions of the lamellae indicated by the colored frames at the {112} side walls. (g)-(i) STEM-Moiré pattern of the selected regions of the lamellae as highlighted by the colored frames at the {110} side walls of NW.

From the Moiré analysis (<112> scan direction) of the NW cross-sections (panels d-i) of each sample we could identify some structural defects. In particular, in the sample with InP nominal thickness of 1 nm some dislocations could be found located at the corners (Fig. 5 (d)). The InP shell is not well developed here and this could be the reason of the presence of structural defects in this region of the sample. In fact, when all InP facets are well developed, as in case of 4 nm, no defects were found at the corners (see panel (e)). However, for thicker InP shell (8 nm) some dislocations appear (see panel (f) and also figure S3 of the supporting information). We have investigated carefully also the {110} side walls of the CDS NWs, as shown in panels (g, h, i). While the samples with InP nominal thickness of 1 and 4 nm do not show any dislocation, the sample with InP nominal thickness of 8 nm shows some dislocations at both the InAs/InP and InP/GaAsSb interfaces and the Moiré pattern shows lattice distortion (panel i). Moreover, in the 8 nm-InP sample, the HR-STEM analysis reveals that the interfaces are not atomically flat and an increased roughness is observed (see supporting information S3 for further detailed TEM images). This roughening may play a role in the relaxation process of the strain of the system.[29]

From the STEM analysis of these lamellae (shown in Fig. S2 of the supporting information) we also found that the actual shell thicknesses are different along the <110> and <112> directions of the NWs, as summarized in Table 1.

**Table 1**: Actual shell thicknesses in the <110> and <112> directions of the NWs.

| Nominal InP thickness (nm) | Measured thickness in <110> direction (nm) | | Measured thickness in <112> direction (nm) | |
|---|---|---|---|---|
| | InP | GaAsSb | InP | GaAsSb |
| 1 | $1.0 \pm 0.1$ | $11.0 \pm 0.9$ | <1.0 | Unknown (interface not visible) |
| 4 | $4.0 \pm 0.9$ | $13.0 \pm 1.0$ | $4.0 \pm 0.9$ | $5.5 \pm 1.2$ |
| 8 | $8.0 \pm 0.5$ | $11 \pm 0.7$ | $9.0 \pm 0.6$ | $6.6 \pm 1$ |

In the sample with 1 nm InP (nominal thickness), the average values for the thickness of InP and GaAsSb along the <110> direction (perpendicular to the InAs side facets) are 1.0 ± 0.1 nm and 11.0 ± 0.9 nm, respectively. On the other hand, along the <112> direction (InAs corners) the InP shell is at least 2 monolayers thinner, as suggested by the lack of a clear HAADF contrast in the image, that makes difficult to measure also the GaAsSb thickness in this direction. In the sample with InP nominal thickness of 4 nm the mean value of GaAsSb shell thickness is 13.0 ± 1.0 nm along the <110> direction and 5.5 ± 1.2 nm along the <112> direction. The InP shell, instead, is uniformly grown along both directions with a thickness of 4.0 ± 0.9 nm in this case. Finally, in the sample with InP nominal thickness of 8 nm we measured GaAsSb and InP shell thickness in the <110> direction of 11 ± 0.7 and 8.0 ± 0.5 nm, respectively. The shell thicknesses of GaAsSb and InP along the <112> direction are 6.6 ± 1 nm and 9.0 ± 0.6 nm, respectively. So the analysis of the shell thickness along the two directions suggests that the growth rate of the GaAsSb shell is higher in the <110> direction as compared to <112> direction, leading to a non-uniform shell thickness. This explains also the development of the low energy <112> facets in the GaAsSb shell.[27] By contrast the InP shell is quite uniform in the two directions when the nominal thickness is higher than 1 nm. For shorter InP growth times, however, the InP shell is well defined only in the <110> direction, suggesting an island-growth mode with preferential nucleation on the InAs side facets. This kind of behavior can be attributed to the different surface energy, surface reconstruction, surface diffusion and nucleation kinetics in the different crystallographic directions.[30] However a deeper analysis of the growth mechanisms is beyond the scope of the present paper.

For the detailed strain analysis at the heterointerfaces, high resolution STEM-HAADF images were acquired and processed with the geometric phase analysis (GPA) method to extract the local components of the strain in the <110> and <112> directions and get strain maps. In general, strain from GPA map is defined as $\varepsilon_{GPA} = (d_{loc} - d_{ref})/d_{ref}$, where $d_{loc}$ is the interplanar spacing of the local part and $d_{ref}$ is the interplanar spacing of reference part which is InAs in our case.[29]

Figure 6 shows the results of our STEM-GPA analysis for the three samples with different InP barrier thickness. Panels (a), (d) and (g) are the STEM images of the three different samples, while panels (b-c), (e-f) and (h-i) are the corresponding GPA maps of $\varepsilon_{xx}$ (i.e. variation of the interplanar spacing in the <112> direction, parallel to the interface) and $\varepsilon_{yy}$ (i.e. variation of the interplanar spacing in the <110> direction, perpendicular to the interface). For the last we report also a line profile across the heterointerfaces (insets).

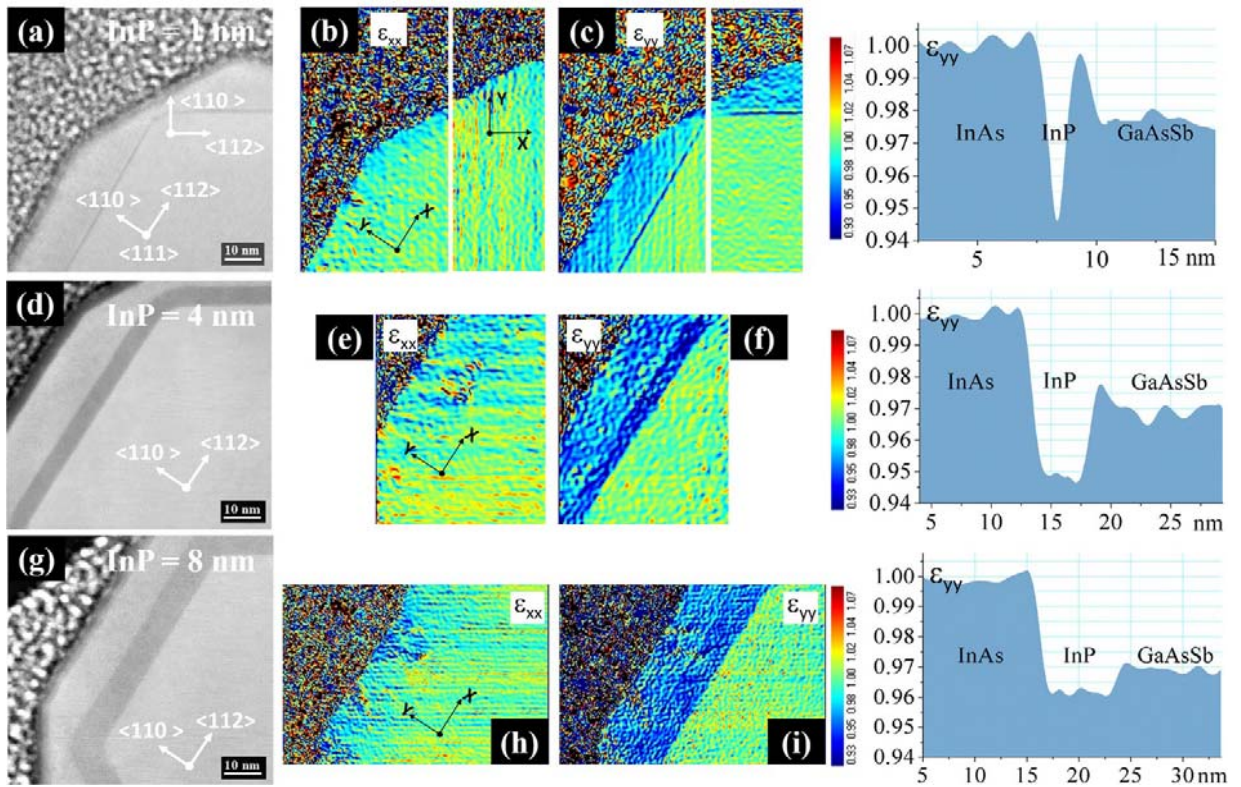

**Figure 6.** HR-STEM images (a, d, g) of the InAs/InP/GaAsSb interface region of the three samples with different InP thickness and corresponding GPA maps of $\varepsilon_{xx}$ (b, e, h) and $\varepsilon_{yy}$ (c, f, i) strain components. The insets are the line profiles of $\varepsilon_{yy}$ across the two interfaces.

Interestingly, no abrupt changes in $\varepsilon_{xx}$ across the interfaces were seen, suggesting that the in-plane lattice parameter of the shell materials is fully strained and adapted to the in-plane lattice parameter of InAs. Only in the 8 nm thick InP sample, an initial relaxation is observed, in agreement with the HRSTEM findings on the presence of dislocations at the interface. On the other hand, from GPA map of $\varepsilon_{yy}$, the two interfaces can be easily identified in all the samples. Indeed, the interplanar spacing perpendicular to the interface varies abruptly, as clearly visible also from the $\varepsilon_{yy}$ line profiles, meaning that the shell lattice in this direction is free to accommodate the strain. The average value of $\varepsilon_{yy}$ for GaAsSb is -3% and for InP is -5%. In order to understand these results, we need to precisely quantify the chemical composition of the two shells, so we performed EDX analysis of these lamellae (see supporting information S4 for the details). In the outer shell we found 40% As and 60% Sb, therefore the chemical composition is $GaAs_{0.4}Sb_{0.6}$. This explains the strain measured from $\varepsilon_{yy}$ line profile, indeed a GaAsSb alloy with such composition is expected to have a lattice parameter in the ZB phase close to 0.59 nm,[12] giving a negative $\varepsilon_{yy\text{-GaSb}}$ as experimentally observed. Concerning the InP shell, from the EDX analysis we found a small As signal also here, but in much smaller concentration (less than 10 atomic %). As already mentioned, the unintentional As incorporation in both shells can be due to the residual As background in the growth chamber after the InAs growth with very high TBAs line pressure. The higher As incorporation during GaSb deposition, compared to the InP deposition can be a consequence of the much higher growth rate of InP (1.4 nm/min) compared with the one of GaAsSb (0.21 nm/min) and to the different Ga and In preferential bonding with As than Sb or P when both group V elements are present in the vapor phase.[31] Moreover it is well known that chemical processes involved in the CBE technique are quite complex and that a large number of possible species and reaction pathways in the vapor phase can complicate the link between precursor flux ratios and the final chemical composition of the grown structure.[32]

Some considerations can be made on the experimental values of $\varepsilon_{yy}$ from the GPA maps. Following the model presented in Ref. 33 for zinc-blende heterostructures grown along the <hhk> direction (in our case: <110>), an epilayer (in our case: shell) pseudomorphically grown on top of a substrate (in our case: core) with a certain lattice mismatch $\Delta a/a$ will assume an out-of-plane parameter determined by its elastic stiffness constants according to a tetragonal distortion process (more details are given in the supporting information). Considering those of $GaAs_{0.4}Sb_{0.6}$ from the literature[34] we get $\Delta d/d = 1.5\ \Delta a/a$. The lattice constant of $GaAs_{0.4}Sb_{0.6}$ is 0.59 nm, so $\Delta a/a = -2.14\ \%$ and we obtain $\varepsilon_{yy} = -3.2\ \%$, which is in good agreement the measured value of $\varepsilon_{yy}$ (about 0.97, being 1 the InAs reference). Similarly, taking the elastic stiffness constants of an InAsP alloy with 10% As, it is expected $\varepsilon_{yy} = 1.7\ \Delta a/a$.[33] Since the lattice mismatch for such $InAs_{0.1}P_{0.9}$ layer grown on InAs is $\Delta a/a = -2.8\ \%$, we obtain $\varepsilon_{yy} = -4.8\ \%$, which is in good agreement with the values of the $\varepsilon_{yy}$ profile at the InAs/InP interface (more details of these calculations are given in the supporting information). It should be noted that $\varepsilon_{yy\text{-}InP}$ in the sample with 8 nm thick InP shell is a bit smaller than the other samples. This is probably related to the partial relaxation that starts to occur in this sample: the out of plane parameter is no more constrained by tetragonal distortion but at the same time it is not yet come back to that of free InP, so that $\varepsilon_{yy}$ takes a value in the range between - 4.8 % (tetragonal distortion) and - 2.8 % (relaxed InP).

From our analysis, we can conclude that InAs/InP/GaAsSb CDS NWs with InP thickness higher than 1 nm and at least up to 4 nm present flat interfaces without dislocations. In particular, the InP shell adopts a lattice parameter coherent with that of the InAs core along the <112> direction (parallel to the interface), while it is elastically compressed along <110> direction (perpendicular to the interface) according to a tetragonal distortion of the lattice. This distortion can accommodate strain energy without misfit dislocations and is a well-known process that occurs when two lattice mismatched materials are grown one on each other, below the critical thickness. Indeed, above a certain thickness the strain energy is too high to be accommodated through a lattice distortion, so the system relaxes more efficiently by producing misfit dislocations.[35] This is consistent with our observation that in InAs/InP/GaAsSb CDS NWs with 8 nm InP some dislocations and interface roughness occur at the interfaces as a consequence of the increased strain field.

CONCLUSIONS

Optimized InAs/InP/GaAsSb core-dual-shell NWs were realized by catalyst-free CBE by varying the growth parameters. We found that the InP shell is not uniformly developed when the nominal thickness is around 1 nm and some dislocations are observed at the corners, where the InP is thinner. For samples with the InP thickness above 1 nm and below 8 nm the InP shell is uniform along all the crystallographic directions and we could not find any dislocation at the heterointerfaces. GPA maps indicate that the strain is accommodated through a tetragonal distortion of the lattice without forming structural defects. On the other hand, when the InP shell is thicker, the interfaces are not flat anymore, but an increased roughness is observed and dislocations start to form. Our study provides useful guidelines for obtaining device-quality InAs/InP/GaAsSb CDS NWs. Moreover, the present approach can be applied to other lattice-mismatched material combinations in order to expand the range of options for device implementation.

## SUPPORTING INFORMATION

Supporting information contains HAADF intensity profiles, high resolution TEM analysis, X-ray energy dispersive analysis, and tetragonal distortion calculation.

## AUTHOR INFORMATION

**Corresponding Author**

*Valentina Zannier (email: valentina.zannier@nano.cnr.it)

- **Author Contributions**

The manuscript was written through contributions of all authors. All authors have given approval to the final version of the manuscript.

- **Funding Sources**

This research activity was partially supported by the SUPERTOP project of QUANTERA ERA-NET Cofound in Quantum Technologies, the FET-OPEN project And QC, the Natural Science Foundation of China (No.51872008), the Beijing Natural Science Foundation (No. Z180014) and the "111" Project under the DB18015 grant.

# Supporting information

# Growth and Strain Relaxation Mechanisms of InAs/InP/GaAsSb Core-Dual-Shell Nanowires

*Omer Arif,[§] Valentina Zannier,[§]\* Ang Li,[#] Francesca Rossi,[∥] Daniele Ercolani,[§] Fabio Beltram,[§] and Lucia Sorba[§]*

[§] NEST, Istituto Nanoscienze-CNR and Scuola Normale Superiore, Piazza San Silvestro 12, I-56127 Pisa, Italy.

[#] Beijing Key Laboratory of Microstructure and Properties of Solids, Beijing University of Technology, 100124 Beijing, China.

[∥] IMEM-CNR, Parco Area delle Scienze 37/A, I-43124 Parma, Italy.

\* Valentina Zannier (email:valentina.zannier@nano.cnr.it)

**HAADF intensity profiles**

Figure S1 shows the high angle annular dark field (HAADF) intensity profiles of (a) InAs/InP and (b) InAs/InP/GaAsSb NWs in cross section. As described in the main text of the article, it is observed that InAs/InP core-shell NWs have six {110} equivalent side facets and the InP shell grew coherently with the InAs core. In case of InAs/InP/GaAsSb core-dual-shell (CDS) NWs, instead, we found that the GaAsSb shell has twelve facets: six belong to the {110} and six belong to the {112} family of planes. So it is clear that development of this kind of facets is related to the GaAsSb shell growth itself and it is not influenced by the InP shell growth, as discussed in the main text.

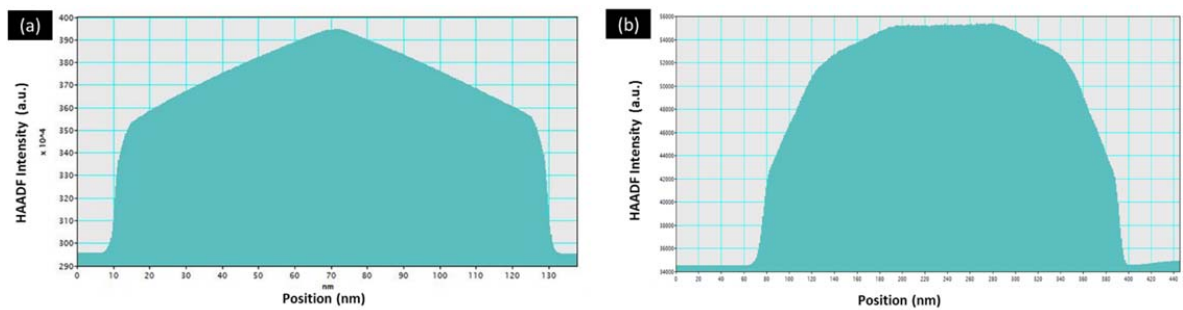

**Figure S1.** STEM-HAADF intensity profiles obtained in cross section on (a) InAs/InP and (b) InAs/InP/GaAsSb nanowires oriented in <112> zone axis. The InAs/InP CS NW has six {110} side facets and InP shell follows the same faceting of the InAs core. The InAs/InP/GaAsSb CDS NW has twelve side facets of the {110} and {112} type.

**High resolution TEM analysis**

Figure S2 shows the high resolution scanning electron microscopy (HR-STEM) images of three InAs/InP/GaAsSb CDS NWs having different InP shell nominal thickness (1, 4 and 8 nm) and the same GaAsSb shell nominal thickness (12 nm). Micrographs were acquired at a <112> facet (panels a, b, c) or a <110> facet (panels d, e, f). These high resolution images helped us to study interfaces and surface roughness at the atomic scale. It is found that the InP shell is uniformly developed only above 1 nm shell thickness, while at 1 nm it is well developed along the <110> facets but it is thinner. The interfaces are atomically flat and without any dislocations for 1 and 4 nm thickness (panels d, e). As long as the InP shell thickness reaches 8 nm, the interfaces between InP/InAs and InP/GaAsSb are not flat anymore and many steps and some dislocations are found (panel f and Figure S3).

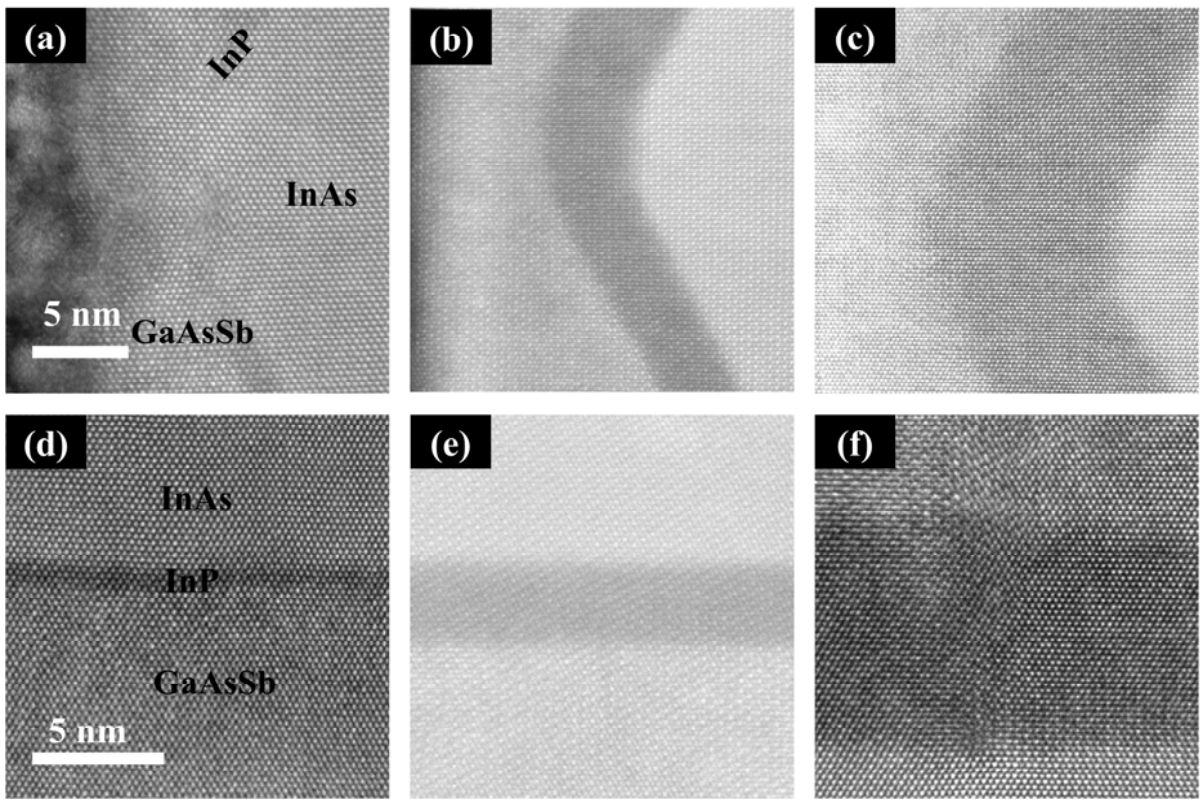

**Figure S2.** HR-STEM images of InAs/InP/GaAsSb CDS NWs at the {112} (a-c) and {110} (d-f) side walls.

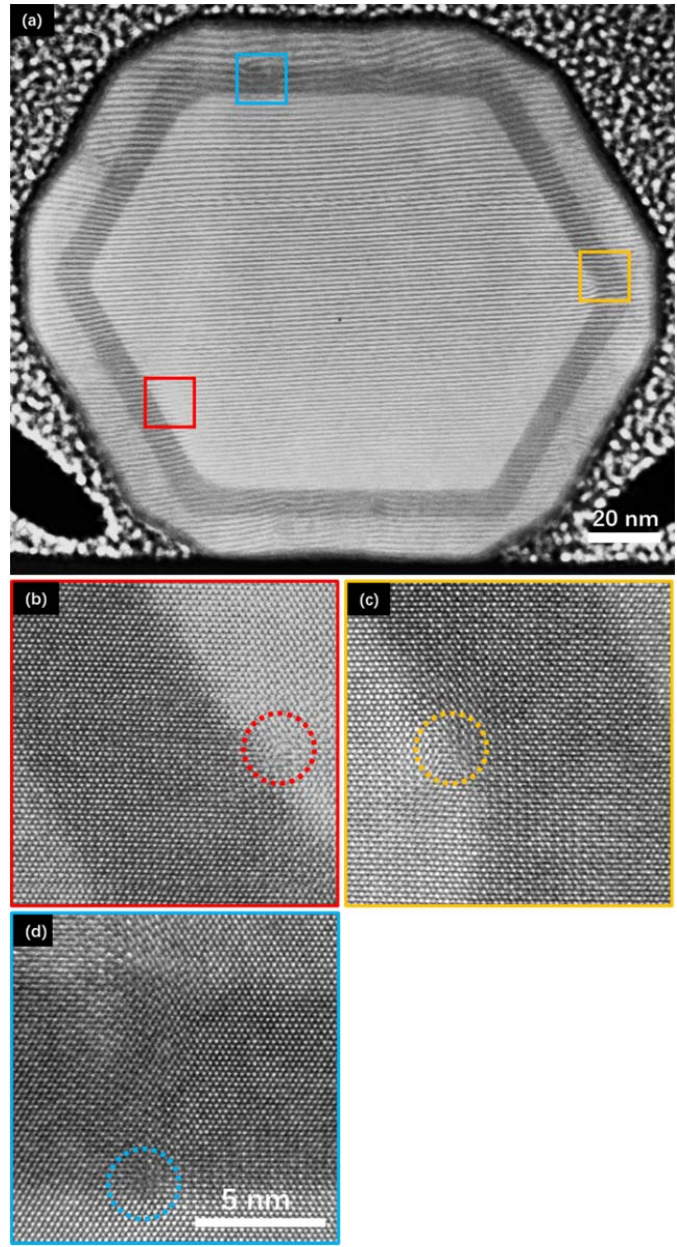

**Figure S3.** STEM analysis of the InAs/InP interface of the sample with 8 nm InP barrier. (a) STEM-Moiré image of entire NW cross section. (b)-(d) HR-STEM images of the defected InAs/InP {110} sidewall interfaces. The colored squares are indicating the corresponding locations and the defects are emphasized by circles.

**Energy dispersive X-ray analysis**

Figure S4 shows the STEM image of the cross-sectional lamellae (a) and the corresponding EDX line profile (b) taken along the arrow indicated in panel (a) of a typical InAs/InP/GaAsSb CDS NW. From the EDX analysis it is found that the outer shell is an alloy instead of a pure GaSb shell. Indeed, from the quantitative analysis we found that the chemical composition is GaAs$_{0.4}$Sb$_{0.6}$. The presence of As inside the GaSb shell is probably related to the use of an high TBAs line pressure for the catalyst-free growth of the InAs core, so that during the following growth of the GaSb shell there is still some residual As in the chamber. Moreover, since the GaSb growth rate at 370°C is very low and the growth time long, the probability of As incorporation is quite high. This problem could be reduced by increasing the GaAsSb shell growth rate, at higher growth temperature, or allowing for As pump out through a much longer growth interruption between the InAs core and the GaSb shell growth.

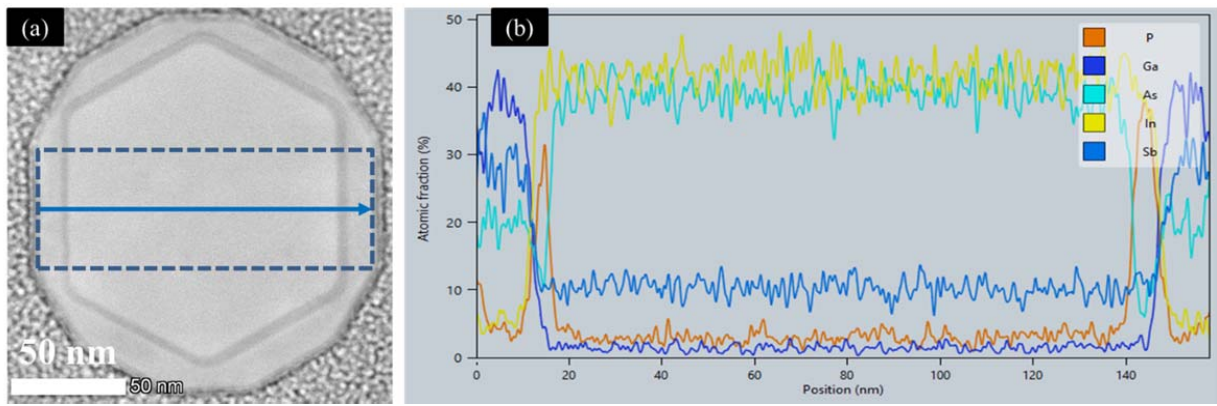

**Figure S4.** Cross-sectional STEM image (a) and elemental line profiles (b) of an InAs/InP/GaAsSb CDS NW.

**Tetragonal distortion**

According to the model presented in Ref. 33 for zinc-blende heterostructures grown along the <hhk> direction (in our case: <110>), an epilayer (in our case: shell) pseudomorphically grown on top of a different material (in our case: InAs core) with a certain lattice mismatch Δa/a will assume an out of plane parameter d determined by the so-called tetragonal distortion. Quantitatively, for {110} interfaces,

$$\frac{\Delta d}{d} = \frac{2c_{11} + 4c_{12}}{c_{11} + c_{12} + 2c_{44}} \frac{\Delta a}{a}$$

Where $c_{11}$, $c_{12}$ and $c_{44}$ are the elastic stiffness constants of the epilayer. In this approximation, for the InP shell over InAs core we can expect Δd/d ($\varepsilon_{yy}$ in GPA notation) =1.7 Δa/a.